\documentclass[prl,twocolumn,amsmath,amssymb,showpacs,showkeys,superscriptaddress]{revtex4}
\usepackage{amsmath}
\usepackage{amssymb}
\usepackage{amsthm}
\usepackage{graphicx}
\usepackage{dcolumn}
\newcommand{\be}{\begin{equation}}
\newcommand{\ee}{\end{equation}}
\newcommand{\ba}{\begin{eqnarray}}
\newcommand{\ea}{\end{eqnarray}}

\begin{document}
\title{Phase diagram of Gaussian-core nematics}
\author{Santi Prestipino}
\email{Santi.Prestipino@unime.it}
\affiliation{Universit\`a degli Studi di Messina, Dipartimento di Fisica, Contrada Papardo, 98166 Messina, Italy}
\author{Franz Saija}
\email{saija@me.cnr.it}
\affiliation{Istituto per i Processi Chimico-Fisici del CNR, Sezione di Messina, Via La Farina 237, 98123 Messina, Italy}
\date{\today}

\begin{abstract} We study a simple model of a nematic liquid crystal made of
parallel ellipsoidal particles interacting via a repulsive Gaussian law.
After identifying the relevant solid phases of the system through a careful
zero-temperature scrutiny of as many as eleven candidate crystal structures,
we determine the melting temperature for various pressure values, also with
the help of exact free energy calculations.
Among the prominent features of this model are pressure-driven reentrant
melting and the stabilization of a columnar phase for intermediate
temperatures.
\end{abstract}



\maketitle
\thispagestyle{empty}

%
%
\section{Introduction}
\setcounter{page}{1}
\renewcommand{\theequation}{1.\arabic{equation}}
Since five decades now, numerical simulation has imposed as an invaluable
tool for the determination of equilibrium statistical properties of
many-particle systems. Despite a long history, however, a precise numerical
evaluation of the Helmholtz free energy of a simple model fluid in its solid
phase has resisted all attacks for many years until, in a remarkable 1984
paper~\cite{Frenkel1}, Frenkel and Ladd showed how a reference Einstein
solid can be used right to this purpose.
Since then, it has become possible to trace a numerically accurate and
complete equilibrium phase diagram for simple-fluid systems by Monte Carlo
simulation methods.
The only real limitation of the Frenkel-Ladd method is the necessity of a
preliminary identification of all relevant solid structures.
Depending on the complexity of the model potential, some structure could be
skipped, neither it does necessarily show up spontaneously in the simulation
due to the effective fragmentation of the system phase space into inescapable
ergodic basins.

In a series of papers~\cite{Saija,Prestipino1,Prestipino2}, we have
employed the Frenkel-Ladd technique in combination with the standard
thermodynamic-integration method in order to trace the phase diagram of
some reference simple-fluid models. In particular, we have provided the
first accurate determination of the phase diagram for the so-called
Gaussian-core model, which is meant to describe dilute solutions of
polymer coils~\cite{Stillinger,Lang}. The thermodynamics of this model
is ruled by the
competition between the fluid and two different, body-centered-cubic
(BCC) and face-centered-cubic (FCC), crystal structures; its peculiar
features are reentrant melting by isothermal compression and, in a
narrow range of temperatures, BCC reentrance in the solid sector.

Following earlier simulational work by Frenkel and collaborators on hard
ellipsoids and spherocylinders~\cite{Frenkel2,Stroobants,Veerman,Bolhuis},
as well as by other authors on hard dumbbells~\cite{Vega},
we aim here to provide another demonstration of the use of simulation for
the description of thermodynamic properties of elongated particles.
Such molecules can exist in a number of partially-ordered mesophases
with long-range orientational order, possibly in combination with one- or
two-dimensional translational order (as in smectic and in columnar liquid
crystals, respectively)~\cite{Pasini,Singh}.
Liquid crystals do also usually give rise to numerous solid phases which,
as a rule, can hardly be anticipated from just a glance at the interaction
potential between the molecules.

Very recently, an interesting liquid-crystal model was introduced by de Miguel
and Martin del Rio \cite{deMiguel} whose phase diagram shows a stable
smectic phase as well as pressure-driven reentrance of the nematic phase.
The model consists of equally-oriented hard ellipsoids that are further
equipped with an attractive spherical well (there is no isotropic phase
in this model since the particles are artificially constrained to stay
parallel to each other, hence the fluid phase is a nematic liquid crystal).
Initially, we thought of this model as an ideal candidate for a complete
reconstruction of the phase diagram.
Unfortunately, the model potential turns out to be not simple enough to
allow for a straightforward identification of the structure of its solid
phase(s) and, in this respect, the original paper is in fact reticent.
We have made an attempt to resolve the solid structure in terms of stretched
cubic lattices but a direct inspection of many equilibrated solid
configurations reveals more complicate, yet periodically repeated patterns.
Probably, this results from a difficult matching between the
optimization requirements of the different pair-potential components,
{\it i.e.}, a cylindrically-symmetric hard-core repulsion and a
spherically-symmetric step-like attraction.

To retain nematic reentrance and, possibly, also the smectic phase, we
have considered a more tractable test case, that is a uniaxial deformation
of the repulsive Gaussian potential, which we expect to provide a model
nematic fluid whose phase diagram can fully be worked out, also in its
solid region.
It can plausibly be argued on symmetry grounds and also expected from the
smoothness of the potential that all solid phases of the model will now
be found within the class of uniaxially-stretched cubic crystals.

The rest of the paper is organized as follows: In Section II, we present
our liquid-crystal model together with a catalogue
of crystal structures that are possibly relevant to it. Next, in Section
III, we outline the numerical methods by which the phase diagram of the
model is being drawn. Results are exposed in Section IV while further
comments and conclusions are deferred to Section V.

%
%
\section{Model}
\renewcommand{\theequation}{2.\arabic{equation}}
\setcounter{equation}{0}
We consider a nematic fluid of $N$ parallel ellipsoids of revolution
whose geometric boundaries are smeared out by a pair interaction
$u$ that smoothly depends on the ratio between the center-to-center
distance $r$ and the ``contact distance'' $\sigma$, which is the
distance of closest approach in case of sharp boundaries.
$\sigma$ is a function of the angle $\theta$ that the ray {\bf r}
joining the two molecular centers forms with the direction
$\hat{\bf z}$ of the axis of revolution. Its closed-form expression
is easily found to be:
\be
\sigma(\theta)=\frac{LD}{\sqrt{L^2\sin^2\theta+D^2\cos^2\theta}}\,,
\label{2-1}
\ee
$D$ and $L$ being the transversal (with respect to $\hat{\bf z}$) and
the longitudinal diameter respectively (we hereafter consider only
the prolate case $L>D$).
For uniaxial particles, the functional dependence of $\sigma$ is
actually on $\cos\theta$, as exemplified by Eq.\,(\ref{2-1}).
We also note that hard ellipsoids do correspond to an interaction
strenght being $+\infty$ for $r<\sigma(\theta)$ and zero otherwise.

For the efficiency of numerical calculation, sufficiently short-range
interaction in all directions is highly desirable and, among smooth
interactions, a good choice is a Gaussian-decaying two-body repulsion,
\be
u(r,\theta)=\epsilon\exp\left\{-\frac{r^2}{\sigma(\theta)^2}\right\}\,,
\label{2-2}
\ee
$\epsilon>0$ being an arbitrary energy scale. Eq.\,(\ref{2-2}) defines
the Gaussian-core nematic (GCN) fluid. It is evident that, upon
increasing the aspect ratio $L/D$, larger and larger system sizes are
needed in order to pull down any rounding-off error that is implicit
{\it e.g.} in the numerical calculation of the total energy.

Another crucial quantity to determine in a simulation is the pressure.
For a $V$-volume system of $N$ parallel ellipsoids in contact with a
heat bath at temperature $T$, the equilibrium pressure $P$ can be calculated
from a virial theorem that generalizes the one valid for a simple fluid.
Let the total potential energy of the system be of the general form
$U=\sum_{i<j}u(|{\bf R}_i-{\bf R}_j|,\cos\theta_{ij})$, where ${\bf R}_i$
is the center-of-mass position of particle $i$ and
$\cos\theta_{ij}=({\bf R}_i-{\bf R}_j)\cdot\hat{\bf z}$.
Upon switching to scaled $V^{-1/3}{\bf R}_i$ coordinates, one readily gets:
\be
P=k_BT\rho-\frac{1}{3V}\left<\sum_{i<j}R_{ij}u'_1(R_{ij},\cos\theta_{ij})\right>\,,
\label{2-3}
\ee
where $u'_1$ is the $u$ derivative with respect to its first argument,
$\rho=N/V$ is the (number) density, and $k_B$ is Boltzmann's constant.
Clearly, $\left<\ldots\right>$ is a canonical-ensemble average.
Upon introducing the $T$- and $\rho$-dependent, two-body distribution
function $g_2({\bf R}_1,{\bf R}_2)=g(|{\bf R}_1-{\bf R}_2|,\cos\theta_{12})$,
the system pressure can also be expressed as
\be
P=k_BT\rho-\frac{\pi}{3}\rho^2\int_{-1}^1{\rm d}\tau\int_0^{+\infty}{\rm d}r\,
r^3g(r,\tau)u'_1(r,\tau)\,.
\label{2-4}
\ee
In particular, for a system of hard ellipsoids the pressure reads:
\be
P=k_BT\rho+\frac{\pi k_BT}{3}\rho^2\int_{-1}^1{\rm d}\tau\,
\sigma(\tau)^3g(\sigma(\tau)^+,\tau)\,.
\label{2-5}
\ee
Anyway, a practical implementation of Eq.\,(\ref{2-4}) or (\ref{2-5}) in
a simulation requires a precise evaluation of the two-argument function
$g$ which, ordinarily, is a difficult task to accomplish with negligible
statistical errors. A much better solution is to switch to the
isothermal-isobaric ensemble, by simulating the system under constant-$T$
and constant-$P$ conditions, see Section III.

As was mentioned in the Introduction, one main inconvenience of
liquid-crystal simulations is the correct identification of the solid
phase(s) of the system, since a plethora of such phases are conceivable
and there is no unfailing criterion for choosing those that are really
relevant to the specific model under investigation. The actual importance
of a given crystal phase can only be judged {\it a posteriori}, after
proving its mechanical stability in a long simulation run and,
ultimately, on the basis of the calculation of its Gibbs free energy,
but nothing can nevertheless ensure that no important phase was skipped.
Besides these vague indications, we adopted a more stringent test in
order to select the phases for which it is worth performing the
numerically-expensive calculation of the free energy.
With specific reference to the model (\ref{2-2}),
we did a comprehensive $T=0$ study of the chemical potential $\mu$
as a function of the pressure for many stretched cubic and hexagonal
phases, in such a way as to identify the stable ground states and leave
out from further consideration all solids with a very large $\mu$
at zero temperature. In fact, it is unlikely that such phases can ever
play a role for the thermodynamics at non-zero temperatures.

For the interaction potential describing the GCN model, we surmise that all
of its stable crystal phases are to be sought among the structures obtained
from the common cubic and hexagonal lattices by a suitable stretching along
a high-symmetry crystal axis, with optimal stretching ratios $\alpha$ that
are probably close to $L/D$. Take {\it e.g.} the case of BCC. We can
stretch it along [001], [110], or [111], this way defining new
BCC001($\alpha$), BCC110($\alpha$), and BCC111($\alpha$) lattices
(the number within parentheses is the stretching ratio; for instance,
BCC001(2) is a BCC crystal whose unit cell has been expanded by a factor
of 2 along $\hat{\bf z}$).
The same can be done with the simple-cubic (SC) and FCC structures.
We further consider hexagonal-close-packed (HCP) and simple-hexagonal
(SH) lattices that are stretched along [111], this way arriving at a total
of eleven potentially relevant crystal phases.

%
%
\section{Method}
\renewcommand{\theequation}{3.\arabic{equation}}
\setcounter{equation}{0}
For fixed $T$ and $P$ values, the most stable of several thermodynamic
phases is the one with lowest chemical potential $\mu$ (Gibbs free
energy per particle).
At $T=0$, only crystal phases are involved in this competition and,
once a list of relevant phases has been compiled,
searching for the optimal one at a given $P$ becomes a simple
computational exercise. An exact property of the Gaussian-core model
(which is the $L/D=1$ limit of the GCN model) is that, on increasing
pressure, the BCC crystal takes over the FCC crystal at
$P^*\equiv PD^3/\epsilon\simeq 0.055$~\cite{Prestipino1}.
Hence, in the GCN model with $L/D>1$ a leading role is naturally
expected for the stretched FCC and BCC crystals.

For an assigned crystal structure, we calculate the $T=0$ chemical
potential $\mu(P)$ of the GCN model for a given pressure $P$ by adjusting
the stretching
ratio $\alpha(P)$ and the density $\rho(P)$ until the minimum of
$(U+PV)/N$ is found. Once the profile of $\mu$ as a function of $P$ is known
for each structure, it is straightforward to draw the $T=0$ phase diagram
for the given $L/D$.

The known thermodynamic behavior at zero temperature provides the general
framework for the further simulational study at non-zero temperatures. In
fact, it is safe to say that the same crystals that are stable at $T=0$ also
give the underlying lattice structure for the stable solid phases at $T>0$.
As we shall see in more detail in the next Section, the only complication
is the existence of three degenerate $T=0$ structures for not too small
pressures, which obliged us to consider each of them as a potentially
relevant low-temperature GCN phase.

We perform a Monte Carlo (MC) simulation of the GCN model with $L/D=3$ in
the isothermal-isobaric ensemble, using the standard Metropolis algorithm
with periodic boundary conditions and the nearest-image convention.
For the solid phase, four different types of lattices are considered,
namely FCC001(3), BCC110(3), BCC111(3), and BCC001(3) (see Section IV).
The number of particles in a given direction is chosen so as
to guarantee a negligible contribution to the interaction energy from
pairs of particles separated by half a simulation-box length in that
direction.
More precisely, our samples consist of $10\times 20\times 8=1600$ particles
in the FCC001(3) phase, of $8\times 24\times 6=1152$ particles
in the fluid and in the solid BCC110(3) phase, of $10\times 12\times 18=2160$
particles in the BCC111(3) phase, and of $12\times 12\times 10=1440$
particles in the BCC001(3) phase.
Considering the large system sizes employed, we made no attempt to
extrapolate our finite-size results to infinity.

At given $T$ and $P$, equilibration of the sample typically took
a few thousand MC sweeps, a sweep consisting of one average attempt per
particle to change its center-of-mass position plus one average attempt
to change the volume by a isotropic rescaling of particle coordinates.
The maximum random displacement of a particle and the maximum volume change
in a trial MC move are adjusted once a sweep during the run so as to keep
the acceptance ratio of moves close to 50\% and 40\%, respectively.
While the above setup is sufficient when simulating a (nematic) fluid
system, it could have harmful consequences on the sampling of a solid
state to operate with a fixed box shape since this would not allow the
system to release the residual stress.
That is why, after a first rough optimization with a fixed box shape,
the equilibrium MC trajectory of a solid state is generated with a
modified (so called constant-stress) Metropolis algorithm which
makes it possible to adjust the length of the various sides of the box
independently from each other (see {\it e.g.} \cite{Stroobants}).
Ordinarily, however, the simulation box will deviate only very little
from its original shape.
When the opposite occurs, this indicates a mechanic instability of the
solid in favor of the fluid, hence it gives a clue as to where melting is
located. We note that MC simulations with a varying box shape are not well
suited for the fluid phase since in this case one side of the box usually
becomes much larger or smaller than the other two, a fact that seriously
prejudicates the reliability of the simulation results.

In order to locate the melting point for a given pressure, we generate
separate sequences of simulation runs, starting from the cold solid
on one side and from the hot fluid on the other side.
The last configuration produced in a given run is taken to be the
first of the next run at a slightly different temperature.
The starting configuration of a ``solid'' chain of runs
was always a perfect crystal with $\alpha=3$ and a density equal to
its $T=0$ value.
Usually, this series of runs is carried on until a sudden change
is observed in the difference between the energies/volumes
of solid and fluid, so as to prevent us from averaging over heterogeneous
thermodynamic states. Thermodynamic averages are computed over
trajectories $10^4$ sweeps long. Much longer trajectories are
constructed for estimating the chemical potential of the fluid
(see below).

Estimating statistical errors is a critical issue whenever
different candidate solid structures so closely compete for thermodynamic
stability. To this aim, we divide the MC trajectory into ten blocks and
estimate the length of the error bars to be twice as large as the standard
deviation of the block averages.
Typically, the relative errors affecting the energy and the volume of the
fluid are found to be very small, a few hundredths percent at the most
(for a solid, they are even smaller).

A more direct clue about the nature of the phase(s) expressed by the
system for intermediate temperatures can be got from a careful monitoring
across the state space of a ``smectic'' order parameter (OP) and of two
different, transversal and longitudinal (with respect to $\hat{\bf z}$)
distribution functions (DFs). The smectic OP is defined as:
\be
\tau(\lambda)=\left|\frac{1}{N}\sum_i\exp\left\{{\rm i}\frac{2\pi D}{\lambda}z_i\right\}\right|\,.
\label{3-1}
\ee
This quantity is able to notice the existence of a layered structure
along $\hat{\bf z}$ in the system, be it solid-like or smectic-like.
In particular, the $\lambda$ at which $\tau$ takes its largest value
gives the nominal distance $\lambda_{\rm max}$ between the layers.
A large value of $\tau$ at $\lambda_{\rm max}$ signals
a strong layering along $z$ with period $\lambda_{\rm max}$.
In order to discriminate between solid and smectic (fluid) layers, we
can rely on the in-plane DF $g_\perp(r_\perp)$, with
$r_\perp={\bf r}-({\bf r}\cdot\hat{\bf z})\hat{\bf z}$, which informs
on how much rapid is the decay of crystal-like spatial correlations in
directions perpendicular to $\hat{\bf z}$.
The persistence of crystal order along $\hat{\bf z}$ is measured through
another DF, $g_\parallel(z)$, which gives similar indications as
$\tau(\lambda)$.
A liquid-like profile of $g_\perp$ along with a sharply peaked $\tau$ or
$g_\parallel$ will be faithful indication of a smectic phase.
Conversely, a sharply peaked $g_\perp$ along with a structureless
$g_\parallel$ will be the imprints of a columnar phase.
Both $g_\perp(r_\perp)$ and $g_\parallel(z)$ are
normalized in such a way as to approach 1 at large distances in case of
fully disordered center-of-mass distributions in the respective directions.
Slight deviations from this asymptotic value may occur as a result of the
variation of box sidelengths during a simulation run.
The two DFs were constructed with a spatial resolution of
$\Delta r_\perp=D/20$ and $\Delta z=L/20$ respectively, and updated every
10 MC sweeps.

We compute the difference in chemical potential between any two
equilibrium states of the system -- say, 1 and 2 -- within the same
phase (or even in different phases, provided they are separated
by a second-order boundary)
by the standard thermodynamic-integration method as adapted to the
isothermal-isobaric ensemble, {\it i.e.}, via the combined use of
the formulas:
\be
\mu(T,P_2)-\mu(T,P_1)=\int_{P_1}^{P_2}{\rm d}P\,v(T,P)
\label{3-2}
\ee and
\be
\frac{\mu(T_2,P)}{T_2}-\frac{\mu(T_1,P)}{T_1}=
-\int_{T_1}^{T_2}{\rm d}T\,\frac{u(T,P)+Pv(T,P)}{T^2}\,.
\label{3-3}
\ee
To prove really useful, however, the above equations require an independent
estimate of $\mu$ for at least one reference state in each phase.
For the fluid, a reference state can be any state characterized by a very
small density (a nearly ideal gas), since then the excess chemical potential
can be estimated accurately through Widom's particle-insertion
method~\cite{Widom}.
The use of this technique for small but finite densities avoids the otherwise
necessary extrapolation to the ideal gas limit as a reference state for
thermodynamic integration. 

In order to calculate the excess Helmholtz free energy of a solid, we resort
to the method proposed by Frenkel and Ladd~\cite{Frenkel1}, based on a
different kind of thermodynamic integration (see Ref.\,\cite{Prestipino2}
for a full description of this method and of its implementation on a
computer). We note that the ellipsoidal symmetry of the GCN particles is
not a complication at all, since the particle axes are frozen and the only
degrees of freedom been left are the centers of mass.
The solid excess Helmholtz free energy is calculated through a series of
$NVT$ simulation runs, {\it i.e.}, for fixed density and temperature.
As far as the density is concerned, its value is chosen in a way such that
complies with the pressure of the low-temperature reference state, that is
the one from which the $NPT$
sequence of runs is started. We wish to emphasize that, thanks to the large
sample sizes employed, the density histogram in a $NPT$ run always turned
out to be sharply peaked, indicating very limited density fluctuations
(hence, negligible ensemble dependence of statistical averages).

%
%
\section{Results}
\renewcommand{\theequation}{4.\arabic{equation}}
\setcounter{equation}{0}
\subsection{Zero-temperature calculations}
For various $L/D$ values in the interval between 1.1 and 3, we have calculated
the $T=0$ chemical potential $\mu(P)$ for our eleven candidate ground states,
with $P$ ranging from 0 to 0.20.
We report in Table 1 the results relative to $L/D=3$
for two values of $P$, 0.05 and 0.20.
An emergent aspect of this Table is the existence of a rich degeneracy that
is only partly a result of the effective identity of crystal structures up
to a dilation.
Take {\it e.g.} the five structures with the minimum $\mu$ (and with the same
density).
While the BCC001 lattice with $\alpha=3$ is obtained from the FCC001 lattice
with $\alpha=3/\sqrt{2}=2.12\ldots$ by a simple $\sqrt{2}$ dilation, there is
no homothety transforming BCC001(3) into BCC110(3) or into BCC111(3)
(in turn equivalent to SC111(1.5)): Points in these three
lattices have different local environments, as can be checked by counting
the $n$th-order neighbors for $n$ up to 4, yet the three stretched
BCC crystals of minimum $\mu$ share the same $U/N$.
Also the pairs FCC110(3), FCC111(3) and SC001(3), SC110(3) consist of
topologically-different degenerate structures.
This fact is an emergent phenomenon whose deep reason remains unclear to us;
it should deal with the dependence of $u$ on the ratio $r/\sigma(\theta)$,
since the same symmetry holds with a polynomial, rather than Gaussian,
dependence.

For the case of $L/D=3$, we show in Fig.\,1 the overall $P$ dependence at
$T=0$ of the chemical potential $\mu$ for the various solids.
The solid with the minimum $\mu$ is either of the type FCC001 (with
$\alpha=3$) or, say, of the type BCC001 (with $\alpha=3$), a fact that
holds true, but with $\alpha=L/D$, for all $1<L/D<3$.
Other solids are definitely ruled out, and the same will probably hold
for $T>0$.
On increasing $L/D$, the transition from a FCC-type to a BCC-type phase
occurs at a lower and lower pressure, whose reduced value is slightly less
than 0.02 for $L/D=3$.

\subsection{Monte Carlo simulation}
In order to investigate the thermodynamic behavior of the GCN model at
non-zero temperatures, we have carried out a number of MC simulation runs
for a GCN system with $L/D=3$, which is the system with the strongest
liquid-crystalline features that we can still manage numerically.

We have effected scans of the phase diagram for six different pressure values,
$P^*=0.01,0.02,0.03,0.05,0.12$, and 0.20.
With all probability, FCC001(3) is the stable system phase only in
a very small pocket of $T$-$P$ plane nearby the origin.
However, we decided not to embark on a free-energy study of the relative
stability of fluid, FCC001(3), and BCC-type phases at such low pressures
since this would require a numerical accuracy that is beyond our capabilities.
To a first approximation, the boundary line between FCC001(3) and, say,
BCC111(3) can be assumed to run at constant pressure. For relating data
obtained at different pressures, we have carried out two further sequences
of MC runs along the isothermal paths for $T^*=0.002$ (solids) and
$T^*=0.015$ (fluid).

The Frenkel-Ladd computation of the excess Helmholtz free energy per
particle $f_{\rm ex}$ confirms that the BCC001(3), BCC110(3), and BCC111(3)
solids are nearly degenerate at low temperature.
We take $T^*=0.002,P^*=0.05$ as a reference state for the calculation of
solid free energies. With the density fixed at $\rho=0.08562D^{-3}$,
in every case corresponding to $P^*=0.05$, we find
$\beta f_{\rm ex}=144.461(2)$, 144.470(2), and 144.453(3),
for the three above solids respectively, implying a weak preference for
the BCC111(3) phase.
Then, using thermodynamic integration along the $T^*=0.002$ isotherm
(see Eq.\,(\ref{3-2})), we have studied the relative stability of the
three solids as a function of pressure, up to $P^*=0.20$. The results,
depicted in Fig.\,2, suggest that BCC111(3) is the stable phase
throughout the low-temperature region, the other solids being very good
solutions anyway with near-optimal chemical potentials.

We then follow the thermal disordering of the BCC-type solids for fixed
pressure (with three cases considered, $P^*=0.05,0.12$, and 0.20) through
sequences of isothermal-isobaric runs, all starting from $T^*=0.002$, with
steps of 0.001. Any such sequence is stopped when the values of potential
energy and specific volume have collapsed onto those of the fluid, thus
informing that the ultimate bounds of solid stability are reached
(usually, a solid can hardly be overheated).
The stability thresholds detected this way are fairly consistent with the
indication coming from the DF profiles which, upon increasing temperature,
will eventually show a fluid-like appearance.
Thermodynamic integration (see Eq.\,(\ref{3-3})) is used to propagate
the calculated $\mu$ for $T^*=0.002$ to higher temperatures.

As far as the (nematic) fluid is concerned, we have first generated a
sequence of $NPT$ simulation runs for $P^*=0.05$, starting from $T^*=0.015$.
At this initial point, the excess chemical potential $\mu_{\rm ex}$ was
estimated by Widom's insertion method, obtaining $\mu_{\rm ex}=0.986(5)$.
It is worth noting that, in a long simulation run of as many as
$5\times 10^4$ MC sweeps at equilibrium, the chemical-potential value
relaxed very soon, with small fluctuations around the average and no
significant drift observed.
Our analysis of the fluid phase is completed by further simulation runs
along the isobaric paths for $P^*=0.12$ and 0.20, for which we did not
have the need to compute the chemical potential again since this could
be deduced from the volume data along the $T^*=0.015$ isotherm.

Chemical-potential results along the three isobars on which we focussed are
reported in Figs.\,3 to 5. As is clear, with increasing temperature the
fluid eventually takes over the solids. Among the solids, the BCC111(3)
phase is the preferred one for any temperature and pressure, although the
chemical potential of the other solid phases is only slightly larger.
On increasing pressure, the melting temperature goes down, like in the
Gaussian-core model. The necessity of a matching with the zero-temperature
melting point for $P=0$ will then imply reentrant melting in the GCN model too.
The maximum error on the melting temperature $T_m$, which we estimate to be
about 0.003 (hence not that small), entirely depends on the limited precision
of the fluid $\mu_{\rm ex}$, which then constitutes a major source of error
on $T_m$.

The only conclusion we can draw from the above chemical-potential study
is that BCC111(3) is the most stable {\em solid} phase of the system
(provided the pressure is not too low). However, a closer look at the
DF profiles obtained from the simulation of BCC111(3) raises some doubts
about the {\em absolute} stability of this phase at intermediate temperatures,
whatever the pressure, calling for a different interpretation of the hitherto
considered as BCC111(3) MC data. Take, for instance, the case
of $P=0.05$. Upon increasing temperature, while $g_\perp$ keeps strongly
peaked all the way to melting, the solid-like oscillations of $g_\parallel$
undergo progressive damping until they are washed out completely, suggesting
a second-order (or very weak first-order at the most) transformation of
BCC111(3) into a {\em columnar phase} before melting. This is illustrated
in Figs.\,6 and 7, where the DFs are plotted for a number of temperatures.
A similar indication is got from the behavior of the smectic OP, see
Fig.\,8, whose highest maximum eventually deflates at practically the
same temperature, $T^*\approx 0.005$, at which the oscillations of
$g_\parallel$ disappear.
Note that no appearance of a columnar phase is seen during the
simulation of either BCC110(3) or BCC001(3), nor in the simulation
of FCC001(3) for $P^*=0.01$.
A slice of the columnar phase is depicted in Fig.\,9 (right panels).
In this phase, columns of stacked particles are arranged side by side,
tightly packed together so as to project a triangular solid on the $x$-$y$
plane. Neighboring columns are not commensurate with each other, as implied
by a completely featureless $g_\parallel$.

The probable reason for the instability of the smectic phase in the GCN
model is the absence of an {\it ad hoc} mechanism for lateral attraction
between the molecules, which is present instead in the model of
Ref.\,\cite{deMiguel}. By the way, hard ellipsoids do not show a smectic
phase either~\cite{Frenkel2}, at variance with (long) hard spherocylinders
where particle geometry alone proves sufficient to stabilize a periodic
modulation of the number density along $\hat{\bf z}$~\cite{Bolhuis}.

Given the compelling evidence of a columnar phase in the GCN model, one
may now ask whether the conclusions drawn from the chemical-potential data
are all flawed.
In particular, the $\mu$ curves that are tagged as BCC111(3) in Figs.\,3
to 5 would be meaningless beyond a certain temperature $T_c<T_m$. In fact
they are not, {\it i.e.}, they retain full validity up to melting since
the (nearly) continuous character of the transition from BCC111(3) to
columnar allows one to safely continuate thermodynamic integration across
the boundary, with the proviso that what previously treated as the BCC111(3)
chemical potential beyond $T_c$ is to be assigned instead to the columnar
phase.

As pressure goes up, the transition from BCC111(3) to columnar takes place
at lower and lower temperatures. In order to exclude that the columnar phase
too, likewise the fluid, will show reentrant behavior at low pressure, we
have simulated the disordering of a BCC111(3) solid also for $P^*=0.02$ and
0.03 (in fact, no reentrance of the columnar phase is observed). Further
points on the melting line for $P=0.01,0.02$, and 0.03 are fixed through
the behavior of $g_\perp$ as a function of temperature.
All in all, the overall GCN phase diagram appears as sketched in Fig.\,10.
This is similar to the phase portrait of the Gaussian-core model, see Fig.\,1
of Ref.\,\cite{Prestipino2}, with the obvious exception of the columnar phase.
There is a small discrepancy between the melting points as located through
free-energy calculations (full dots in Fig.\,10) and those assessed from the
evolution of $g_\perp$ (open dots). In our opinion, this would mostly be
attributed to the statistical error associated with the $\mu_{\rm ex}$ of the
fluid in its reference state. Notwithstanding their limited precision, however,
free-energy calculations are all but useless in identifying the structure of
the solid phase.
In conclusion, although some aspects of the equilibrium behavior of the GCN
model remain still uncertain, especially with regard to the exact location
of the solid-solid transition at low pressure, we are confident that the main
features of the GCN phase diagram are correctly accounted for by Fig.\,10.

Summing up, there are at least two conceivable and mutually exclusive paths
for the thermal disordering of a liquid-crystal solid (aside from a direct
transformation of it into a nematic phase). One is through the formation of a
smectic phase, which eventually transforms into a nematic fluid.
A second possibility is a more gradual release of crystalline order
by the appearance of a columnar phase as intermediate stage between the
solid and the nematic phase.
Our study showed that it is this second scenario that occurs in the GCN model,
with no evidence whatsoever of a smectic phase.

%
%
\section{Conclusions}
We have introduced a liquid-crystal model of softly-repulsive parallel
ellipsoids, named the Gaussian-core nematic (GCN) model, aiming at a
complete characterization of its phase behavior, including the
solid sector. This requires a preliminary identification of all relevant
solid structures, which is generally a far-from-trivial task to be
accomplished for model liquid crystals~\cite{Pfleiderer}.
Through a careful scrutiny of as many as eleven uniaxially-deformed cubic
and hexagonal phases, we obtained a thorough description of the $T=0$
equilibrium phase portrait of the GCN model, identifying its ground state
at any given pressure. In doing so, we discovered a rich and absolutely
unexpected structural degeneracy, which is only lifted by going to $T>0$.
At low temperature, and for not too low pressures, our free-energy
calculations indicate that a GCN system with an aspect ratio of 3
is found in just one solid phase, {\it i.e.}, a stretched BCC solid
with the molecules oriented along [111]. Only near zero pressure,
the stable phase becomes a stretched FCC solid. With increasing
temperature, the BCC-type solid first undergoes a weak transition
into a columnar phase, which still retains partial crystalline order,
before melting completely into the nematic fluid.

It is worth emphasizing that our interest in the GCN model is purely
theoretical, hard-core ellipsoids providing a more physically realistic
model liquid crystal. One could even argue that a Gaussian repulsion is
highly irrealistic for a liquid crystal. In real atomic systems,
superposition of particle cores is strongly obstructed, whence the
consideration of hard-core or steep inverse-power repulsion in the more
popular models.
However, unless the system density is very high, higher than considered
in our study, repulsive Gaussian particles would effectively be blind to an
inner hard core, which thus may or may not exist, as evidenced {\it e.g.}
in the snapshots of Fig.\,9 where particles appear well spaced out.

The GCN model is a ``deformation'' of Stillinger's Gaussian-core model,
well known for exhibiting a reentrant-melting transition.
Various instances of reentrant behavior are also known for
nematics~\cite{Cladis} and indeed one of the original motivations for the
present work was searching for a new kind of reentrance,
{\it i.e.}, re-appearance of a more disordered phase with increasing pressure.
With this study, we provide yet another example of reentrant behavior in a
model nematic: While this is nothing but the analog of fluid-phase reentrance
in the Gaussian-core model, the absolute novelty of our findings is in the
nature of the intermediate phase, this being surprisingly columnar in a range
of pressures rather than genuinely solid.
%
%

\newpage
%
%
\begin{table}
\caption{GCN model for $L/D=3$: $T=0$ chemical potential $\mu(P)$ for eleven
different solids and two values of $P^*$, 0.05 and 0.20.
$N_x,N_y,N_z$ are the number of lattice points along the three spatial
directions, $\rho=N_xN_yN_z/V$ is the density, and $\alpha$ is
the stretching ratio (for the SH111 lattice, $\alpha$ is the so-called
$c/a$ ratio).
$N_x,N_y,N_z$ have been chosen so large that the rounding-off error on
the total potential energy per particle, $U/N$, due to the finite lattice
size is negligible.
The numerical precision on $\rho$ and $\alpha$ is of one unit on the last
decimal digit.
Looking at the Table, the most stable structures at both pressures are
five degenerate crystals, actually belonging to three distinct types
which are exemplified by BCC001(3) (equivalent to FCC001(2.12) up to
a dilation), BCC110(3), and BCC111(3) (equivalent to SC111(1.5)) --
within brackets is the value of $\alpha$.\label{tab1}
}
\begin{tabular*}{\columnwidth}[c]{@{\extracolsep{\fill}}|r|r|r|r|r||r|r|r|}
\hline
crystal & $N_x,N_y,N_z$ & $\rho(0.05)$ & $\alpha(0.05)$ & $\mu(0.05)$ & $\rho(0.20)$ & $\alpha(0.20)$ & $\mu(0.20)$ \\
\hline\hline
FCC001 & 10,20,10 & 0.086 & 2.12 & 0.855724 & 0.157 & 2.12 & 2.093695 \\
\hline
BCC001 & 14,14,10 & 0.086 & 3.00 & 0.855724 & 0.157 & 3.00 & 2.093695 \\
\hline
SC001 & 20,20,8 & 0.086 & 3.00 & 0.881586 & 0.158 & 3.00 & 2.105241 \\
\hline
FCC110 & 16,12,12 & 0.086 & 3.00 & 0.856391 & 0.157 & 3.00 & 2.094368 \\
\hline
BCC110 & 10,28,8 & 0.086 & 3.00 & 0.855724 & 0.157 & 3.00 & 2.093695 \\
\hline
SC110 & 14,18,10 & 0.086 & 3.00 & 0.881586 & 0.158 & 3.00 & 2.105241 \\
\hline
FCC111 & 16,18,9 & 0.086 & 3.00 & 0.856391 & 0.157 & 3.00 & 2.094368 \\
\hline
BCC111 & 12,12,18 & 0.086 & 3.00 & 0.855724 & 0.157 & 3.00 & 2.093695 \\
\hline
SC111 & 12,12,18 & 0.086 & 1.50 & 0.855724 & 0.157 & 1.50 & 2.093695 \\
\hline
HCP111 & 18,20,10 & 0.086 & 3.00 & 0.856429 & 0.157 & 3.02 & 2.094474 \\
\hline
SH111 & 18,20,9 & 0.086 & 2.75 & 0.870014 & 0.158 & 2.69 & 2.099565 \\
\hline
\end{tabular*}
\end{table}

\newpage
%
%
\begin{figure}
\includegraphics[width=8cm,angle=0]{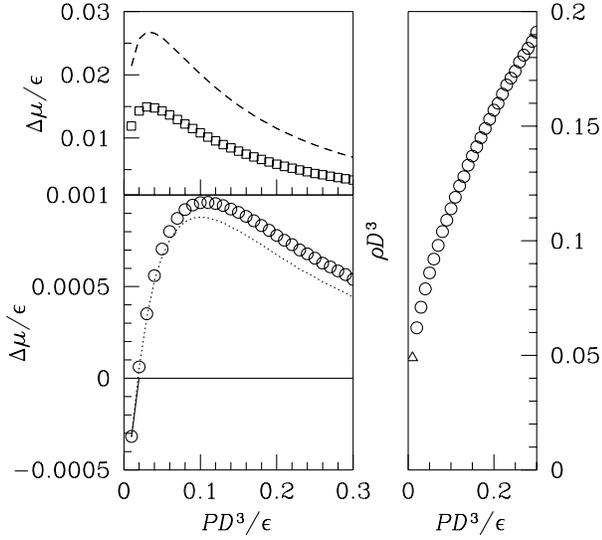}
\caption{$T=0$ equilibrium behavior of the GCN model with $L/D=3$. Left: $T=0$ chemical
potential $\mu(P^*)$ of various crystals relative to BCC110(3), which thus
serves as the zero or reference level. The reduced pressure $P^*$ is
incremented by steps of 0.01. Note that, for all $P$, the five crystals
FCC001(2.12), BCC001(3), BCC110(3), BCC111(3), and SC111(1.5) are degenerate
($\Delta\mu=0$).
Other data points are for FCC001 (continuous line; $\alpha=3$ for $P^*=0.01$,
being $\alpha=2.12$ otherwise), FCC110(3) and FCC111(3) (dotted line),
HCP111 (open dots), SH111 (open squares), SC001(3) and SC110(3) (dashed line).
Right: Resulting equation of state in the pressure range from 0 to 0.30.
FCC001(3) (open triangle) is stable at very low pressure, up to slightly
less than 0.02, while FCC001(2.12), BCC001(3), etc. (open dots) prevail
for higher pressures.\label{fig1}
}
\end{figure}

%
%
\begin{figure}
\includegraphics[width=9cm,angle=0]{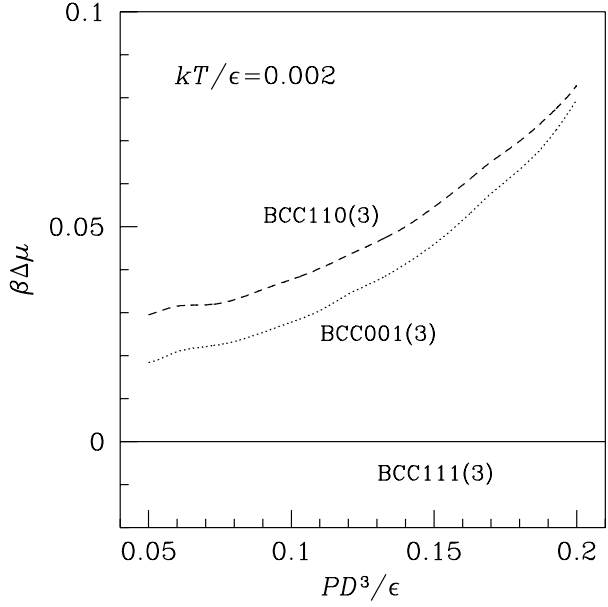}
\caption{GCN model with $L/D=3$, chemical-potential results for $T^*=0.002$.
In the picture, we plot the reduced chemical potential of the three $T=0$
degenerate structures that exist for not too low pressure, taking BCC111(3)
for reference. The latter phase gives the most stable solid for any $P$ in
the range from 0.05 to 0.20 (and, most likely, even further).
The $\mu$ curves are obtained by thermodynamic integration of volume MC data,
using as initial conditions those specified by the Frenkel-Ladd calculations
that were carried out at $P^*=0.05$. Though the reported $\mu$ values for
the BCC-type solids are very close to each other and also affected by
some numerical noise, the higher stability of BCC111(3) cannot be truly
called into question -- a regular pattern is clearly seen behind each curve.\label{fig2}
}
\end{figure}

%
%
\begin{figure}
\includegraphics[width=8cm,angle=0]{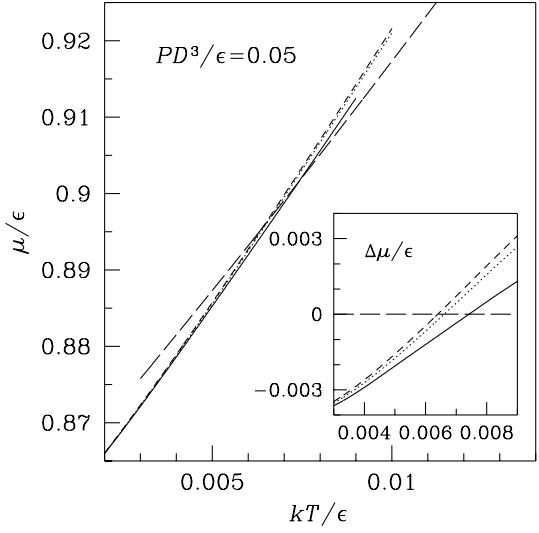}
\caption{GCN model with $L/D=3$, chemical-potential results for $P^*=0.05$:
Chemical potential of the fluid phase (dotted line) as compared with
those of the competing solid phases for that pressure (BCC001(3),
long-dashed line; BCC110(3), dashed line; BCC111(3), continuous line).
While the BCC111(3) solid is stable at low temperature, the fluid phase
overcomes it in stability for higher temperatures. This is more clearly
seen in the inset, where chemical-potential differences are reported,
taking the fluid $\mu$ for reference. The melting temperature for
$P^*=0.05$, which is where the continuous line crosses zero, is
estimated to be $T^*\simeq 0.0073$.\label{fig3}
}
\end{figure}

%
%
\begin{figure}
\includegraphics[width=8cm,angle=0]{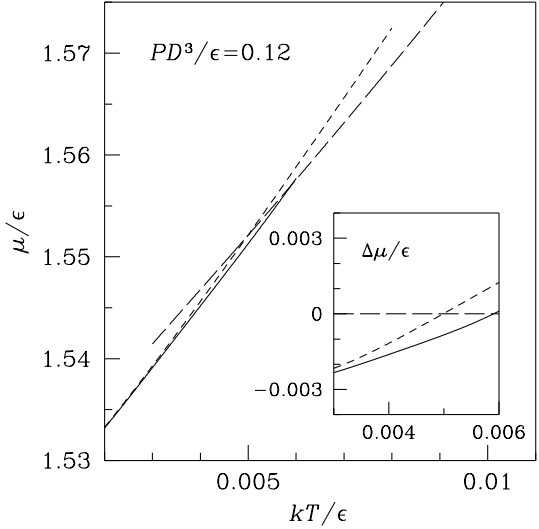}
\caption{GCN model with $L/D=3$, chemical-potential results for $P^*=0.12$.
Same notation as in Fig.\,3, except for the absence of data for BCC001(3),
which were not computed. Despite this, a look at the results in Figs.\,2
and 3 give us confidence that the chemical potential of BCC001(3) will be
closer to that of BCC110(3) than is for $P^*=0.05$.\label{fig4}
}
\end{figure}

%
%
\begin{figure}
\includegraphics[width=8cm,angle=0]{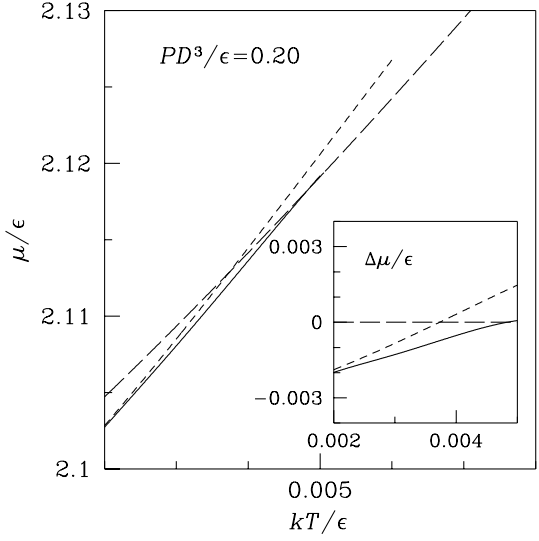}
\caption{GCN model with $L/D=3$, chemical-potential results for $P^*=0.20$.
Same notation as in Figs.\,3 and 4.\label{fig5}
}
\end{figure}

%
%
\begin{figure}
\includegraphics[width=8cm,angle=0]{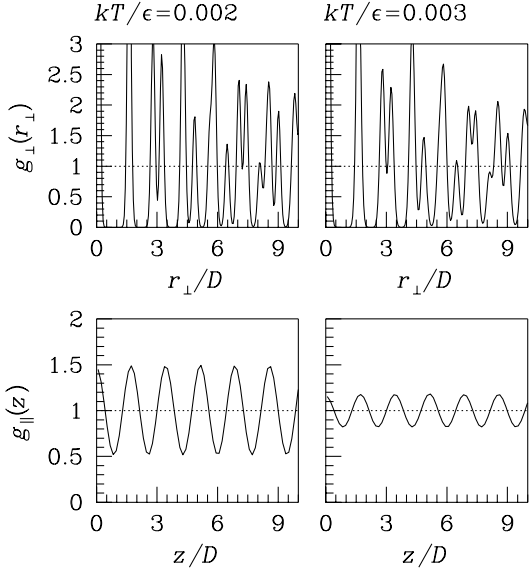}
\caption{GCN model with $L/D=3$, distribution functions of BCC111(3) for $P^*=0.05$.
Left: $T^*=0.002$. Right: $T^*=0.003$.
The strenght of crystalline order along $\hat{\bf z}$, as measured
by the amplitude of $g_\parallel$ oscillations, reduces with
increasing temperature, until complete disorder is left above
$T^*\simeq 0.005$ (see next Fig.\,7). Considering that the crystallinity
within the $x$-$y$ plane persists well beyond $T^*=0.005$ (the spatial
modulation of $g_\perp$ remains solid-like beyond this temperature and
up to melting), we conclude that the GCN system is found in a columnar
phase for $0.005<T<T_m$.\label{fig6}
}
\end{figure}

%
%
\begin{figure}
\includegraphics[width=8cm,angle=0]{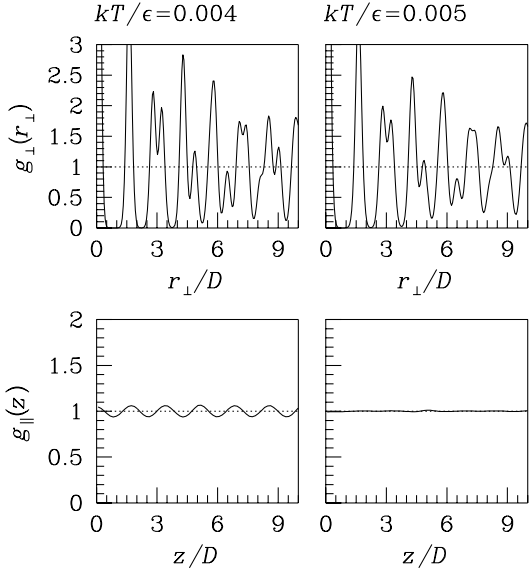}
\caption{GCN model with $L/D=3$, distribution functions of BCC111(3) for $P^*=0.05$.
Left: $T^*=0.004$. Right: $T^*=0.005$.\label{fig7}
}
\end{figure}

%
%
\begin{figure}
\includegraphics[width=8cm,angle=0]{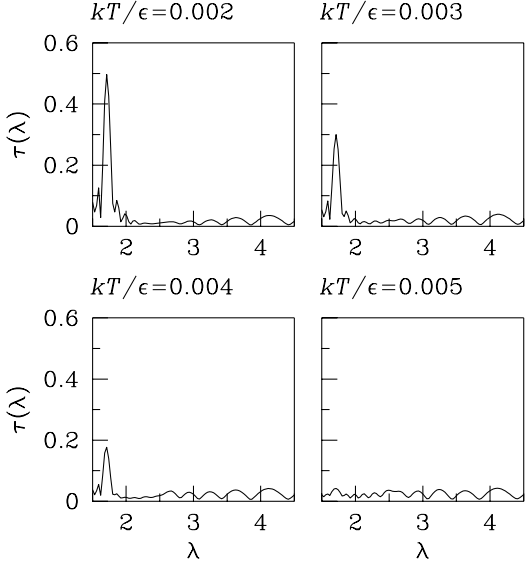}
\caption{GCN model with $L/D=3$, smectic order parameter $\tau(\lambda)$ of BCC111(3)
for $P^*=0.05$. The behavior of $\tau(\lambda)$ faithfully reproduces that
seen for $g_\parallel(z)$ (cf. Figs.\,6 and 7): The deflating of the
highest $\tau(\lambda)$ maximum with increasing temperature closely
follows the thermal damping of $g_\parallel(z)$ oscillations.\label{fig8}
}
\end{figure}

%
%
\begin{figure}
\includegraphics[width=8cm,angle=0]{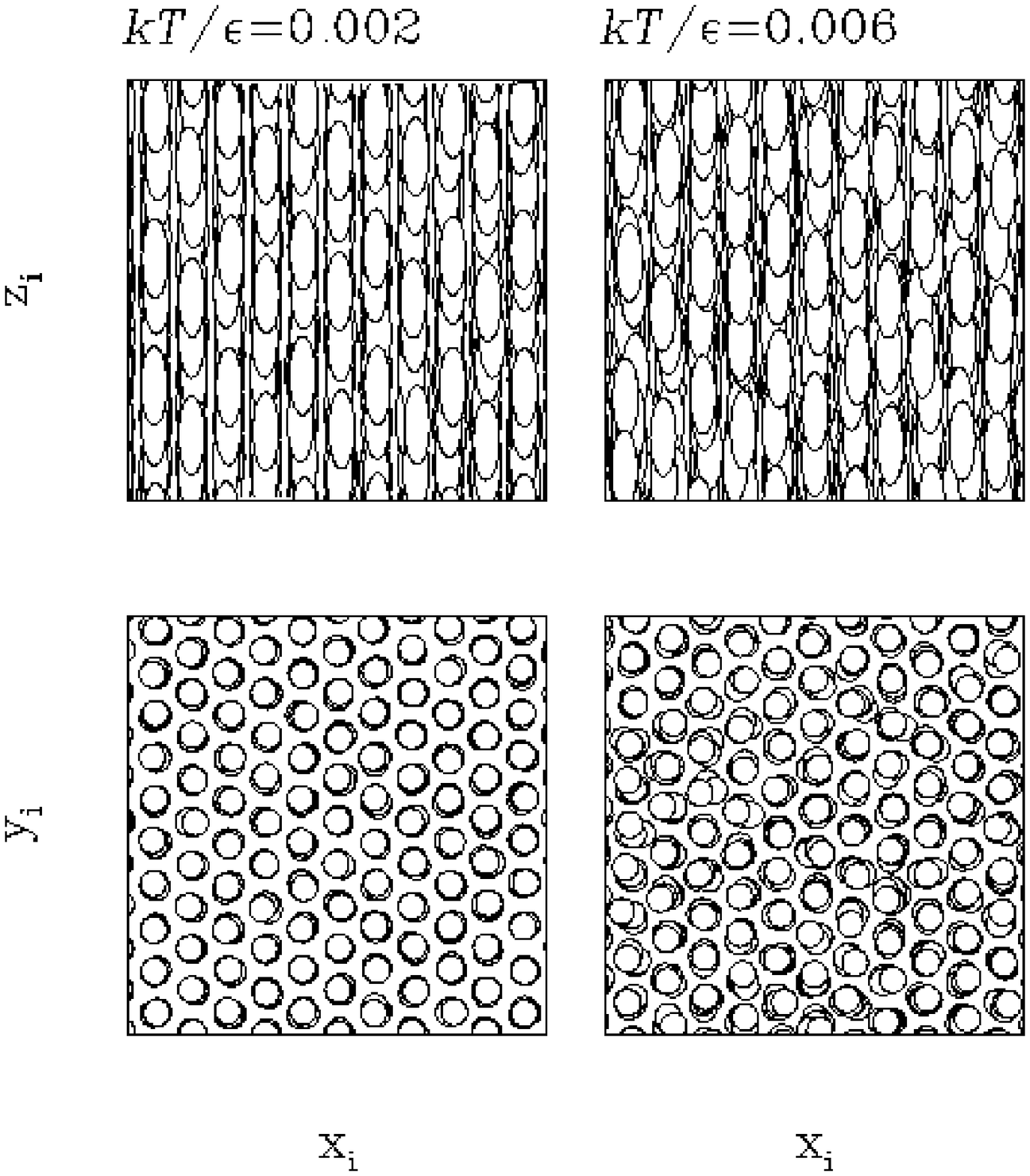}
\caption{GCN model with $L/D=3$, some snapshots of the particle configuration
taken at low temperature ($T^*=0.002$, BCC111(3) solid phase) and at
intermediate temperature ($T^*=0.006$, columnar phase). The reduced
pressure is $P^*=0.05$ in both cases. Above: side view, {\it i.e.},
projection of particle coordinates onto the $x$-$z$ plane. Below: top
view, {\it i.e.}, projection of particle coordinates onto the $x$-$y$
plane. For clarity, in spite of their mutual interaction being soft,
the particles are given sharp ellipsoidal boundaries, corresponding to a
unitary short axis ($D$) and a long axis of $L=3D$. While the crystalline
order along $z$ is lost already at $T^*=0.005$ (hence, it is there in
the top-left panel while it is absent in the top-right panel), the
triangular order within the $x$-$y$ plane is maintained up to the
melting temperature (here, $T_m\simeq 0.0073$).\label{fig9}
}
\end{figure}

%
%
\begin{figure}
\includegraphics[width=8cm,angle=0]{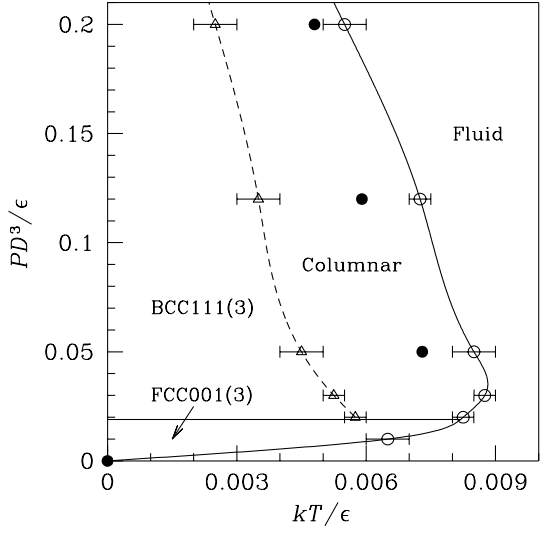}
\caption{GCN model with $L/D=3$, sketch of the phase diagram on the $T$-$P$ plane.
The full dots mark the location of the melting transition as extracted
from our free-energy calculations. Open symbols refer instead to the
transition thresholds as given by a visual inspection of the DF profiles.
Though the latter melting-point estimates are more easily obtained than
the former, the free-energy study was essential to identify the correct
solid structure of the GCN model at not too low pressure.
To help the eye, tentative phase boundaries are drawn as continuous
({\it i.e.}, first-order) and dashed (nearly second-order) lines through
the transition points. In the low-pressure region, the solid-solid boundary
is highly hypothetical since we have no data there.\label{fig10}
}
\end{figure}
\end{document}